\documentclass[letter,onecolumn,draftcls]{IEEEtran}
\usepackage{cite}
\ifCLASSINFOpdf
   \usepackage[pdftex]{graphicx}
   \usepackage{epstopdf}
\else
   \usepackage[dvips]{graphicx}
   \DeclareGraphicsExtensions{.eps}
\fi
\usepackage[cmex10]{amsmath}
\usepackage{array}

\usepackage[lofdepth,lotdepth]{subfig}
\usepackage{verbatim}
\usepackage{pseudocode}
\usepackage{fancybox,amssymb}

\begin{document}
\include{fonts}
\title {Wideband Waveform Design for Robust Target Detection}

\author{Ashkan~Panahi$^*$, Marie~Str\"{o}m$^{*\dagger}$, Mats~Viberg$^*$\\
*\ Signal Processing Group, Signals and Systems Department, Chalmers University, Gothenburg, Sweden\\
$\dagger$\ Saab EDS, Gothenburg, Sweden\\
Email:{ \it\{ashkanp,marie.strom,viberg\}@chalmers.se}}

%\author{\IEEEauthorblockN{Marie~Str\"{o}m\IEEEauthorrefmark{1}\IEEEauthorrefmark{2},
%Ashkan~Panahi\IEEEauthorrefmark{1},
%Mats~Viberg\IEEEauthorrefmark{1} and
%Kent~Falk\IEEEauthorrefmark{2}}
%\IEEEauthorblockA{\IEEEauthorrefmark{1}Signal Processing Group, Signals and Systems Department, Chalmers University, G\"{o}teborg, Sweden %\\ \{{\it maristro, ashkanp, viberg}\}@chalmers.se}
%\IEEEauthorblockA{\IEEEauthorrefmark{2}Saab EDS, Solhusgatan 10, Kalleb\"{a}cks Teknikpark, G\"{o}teborg, Sweden}
%}

%\author{\IEEEauthorblockN{Marie Str\"om}
%\IEEEauthorblockA{Department of Signals and Systems\\
%Chalmers University of Technology\\
%Gothenburg, Sweden \\
%Email: marie.strom@chalmers.se}
%\and
%\IEEEauthorblockN{Mats Viberg}
%\IEEEauthorblockA{Department of Signals and Systems\\
%Chalmers University of Technology\\
%Gothenburg, Sweden \\
%Email: viberg@chalmers.se}
%}

\maketitle

\begin{abstract}
Future radar systems are expected to use waveforms of a high bandwidth, where the main advantage is an improved range resolution. In this paper, a technique to design robust wideband waveforms for a Multiple-Input-Single-Output system is developed. The context is optimal detection of a single object with partially unknown parameters. The waveforms are robust in the sense that, for a single transmission, detection capability is maintained over an interval of time-delay and time-scaling (Doppler) parameters. A solution framework is derived, approximated, and formulated as an optimization by means of basis expansion. In terms of probabilities of detection and false alarm, numerical evaluation shows the efficiency of the proposed method when compared with a Linear Frequency Modulated signal and a Gaussian pulse.
\end{abstract}

\section{Introduction}
\label{sec:introduction}
Signal processing techniques for radar systems have to a great extent focused on narrowband signals. At this time, signal generators are able to synthesize arbitrary signals with a bandwidth of the order of GHz~\cite{Han02,Zhu09,Wentzloff06}. On one hand, this provides interesting new opportunities, e.g., a wideband signal achieves an increased range resolution compared to its narrowband counterpart\cite{taylor94,Khan05,Hussain98,Weiss94}. On the other hand, the simplifying narrowband assumption, where velocity is approximated as a frequency shift, is not valid\cite{Weiss94}. This does not only complicate the design, but also disqualifies traditional detection techniques, as estimation of time-delay and Doppler-shift can not be separated in time and frequency. It should also be considered that, for some applications, maintaining a low computational complexity is crucial, which is naturally against the original desire of super-resolution. Accordingly, this work is devoted to provide an adaptive detection scheme, which establishes an arbitrary trade-off between complexity and resolution.

The above concern is related to other contributions in the area of waveform design. For example, design of wideband ambiguity functions with narrow peaks for Orthogonal-Frequency-Division-Multiplexing signals is considered in~\cite{Sen10,Sen09}. In~\cite{he12} various techniques for designing narrowband or wideband waveforms are discussed. Interesting analysis of wideband radar  systems from various perspectives are also found in~\cite{Lush91,Yazici06,Antonio05}. However, these studies mostly focus on designing ambiguity functions with a narrow peak neglecting complexity. It is easy to see that applying these methods to the discussed topic leads to a large number of transmissions, due to lack of shift-invariance properties, as well as a large set of detection filters.
%However, one finds them irrelevant to the case of interest herein for the following reason. Note that in the multiple-input case, the shift-invariance properties are lost and thus super-resolution may only be uniformly obtained by multiple transmissions with waveforms steered to small parameter sets. Then, the number of necessary pulse transmissions in a radar sweep proportionally increases with decreasing the peak width. Furthermore, a narrow peak needs a larger set of filters to detect.

To combat complexity, this work proposes a waveform design method that maintains a necessary mainlobe width in the ambiguity function, thus decreasing the overall number of pulse transmissions and receive filters. Although this leads to detections of a restricted resolution, it is easy to adapt a secondary super-resolution detection procedure~\cite{borison92,cuomo99,moore95} based on the original estimates. This provides a highly flexible design with low complexity.

%The idea is to follow the method of detection by filter banks and to adapt a double-stage scheme, where in the first stage, low-resolution detections are obtained according to a computational budget. In the next stage, if required, a secondary improved detection by a super-resolution technique~\cite{borison92,cuomo99,moore95} is performed based on the low-resolution results. Note that, since a super-resolution technique is applied to a restricted area, obtained from the first stage, its performance, in terms of speed and accuracy, is expected to increase. The emphasis of this work is on the first stage, and further investigation on highly accurate estimates is postponed to future work.

Here, focuses is on obtaining original estimates of a restricted resolution. This is carried out by considering relatively wide parameter ranges and designing corresponding waveforms, for which reliable detection, in the entire mesh, is ensured after filtering matched to a nominal value. This is not straightforward as wideband signals may lead to a focused ambiguity function, where off-grid targets are easily missed. A matched filter design is selected to ensure a good performance in presence of noise, assuming nominal parameter values. The main contribution of this work is to formulate this idea, and to provide waveforms that guarantee robust detection in a desired interval of target parameters.
In what follows, a statistical framework for detecting a single target is developed, and for which expressions are simplified by approximation to obtain a tractable design.
%The design is formulated as an optimization by means of basis expansion, for which we develop a method to solve. The resulting performance, in terms of probabilities of detection and false alarm, is investigated by means of numerical evaluation.

\section{Problem Formulation}
Consider a bistatic radar system that, on the transmitter side, employs $M$ waveform generators each connected to an antenna element. The receiver side comprises one antenna element connected to a filter bank. Each generator samples a baseband signal composed of a set of $N$ basis functions $\psi_{m,n}(t)$, where $m$ and $n$ are the antenna and basis label, respectively. In other words,
\begin{equation}
\label{eq:basisexpansion}
\tilde{x}_m(t) = \sum_{n=1}^N s_{m,n}\psi_{m,n}(t).
\end{equation}
where $\tilde{x}_m$ is the waveform at the $m$th signal generator, and $s_{m,n}$ is a complex scalar coefficient. The received signal is a mixture of the reflected transmitted waveforms, and can be expressed, for a point target, as
\begin{equation}
\label{eq:signal}
y(t) = \sigma_t \sum_{m=1}^M x_m(\mu(t-\tau_{m}(\phi)-\tau)) + n(t).
\end{equation}
where $\sigma_t$ is the object's reflection coefficient, $\tau$ denotes the time-delay from the zero-phase sensor to the receiver, and $\mu$ is the time-scaling related to the velocity of an object\cite{Hussain98,Weiss94}. Furthermore, $\tau_{m}(\phi)$ is given by the inter-element spacing and the spatial direction, $\phi$, towards the object. This direction, azimuth and/or elevation, is assumed to be known. If this is not the case, a beamforming technique, see, e.g., \cite{gershman1999,li2006,vorobyov2003}, is necessary. In~\eqref{eq:signal}, $x_m(t) = \tilde{x}_m(t)e^{j\omega_ct}$ is centered around the system's carrier frequency $f_c=\omega_c/2\pi$ and $n(t)$ is a white Gaussian noise.

At the receiver side the down-converted signal, $ \tilde{y}(t) = y(t)e^{-j\omega_ct}$, is passed through a filter bank, i.e.,
\begin{equation}
\label{eq:correlation}
r(\tau,\tau',\mu,\mu') = \int h^\ast (t;\tau',\mu') \tilde{y}(t)dt.
\end{equation}
Here, $(\cdot)^\ast$ denotes the complex conjugate. If a so-called matched filter structure\cite{Skolnik,Skolnik81} is employed, the correlating filters, $h(t;\tau',\mu')$, are equal to the received signal calculated from their corresponding transmission model.

To ease calculations, assume that the basis kernels, $\psi_{m,n}(t)$, are Gaussian. This particular choice of functions results in that~\eqref{eq:correlation} can be analytically calculated to
\begin{equation}
\begin{aligned}
\label{eq:output_matchedfilter_2}
&r(\tau_0,\mu,\mu') = \sum_{\substack{m,m'\in \mathcal{M} \\ n,n'\in \mathcal{N} }} \frac{\sigma_t\sqrt{\mu\mu'} s_{m,n}s_{m',n'}e^{-j\omega_c\mu'(\tau_0 + \tau_{m,m'})}}{\sqrt{2\pi(\mu^2\sigma_{m',n'}^2 +\mu'^2\sigma_{m,n}^2)}} \cdot\\
&e^{-\frac{(\mu'\mu_{m,n}- \mu \mu_{m',n'} + \mu \mu'(\tau_0 + \tau_{m,m'}))^2}{2(\mu^2\sigma_{m',n'}^2 +\mu'^2\sigma_{m,n}^2)}}e^{\omega_c(\mu-\mu')(j\mu_G -\frac{\omega_c}{2}(\mu-\mu')\sigma_G^2)},
\end{aligned}
\end{equation}
where  $\mathcal{M} = \{1\dots M\}$, $\mathcal{N} = \{1\dots N\}$, $\sigma_{m,n}^2$ and $\mu_{m,n}$ correspond to the variance and the mean of the $n$th basis kernel sampled by the $m$th signal generator, respectively, $\tau_0 = \tau-\tau'$, $\tau_{m,m'} =\tau_{m}(\phi)-\tau_{m'}(\phi)$, and
\begin{equation}
\begin{aligned}
& \mu_G = \frac{\mu\mu_{m,n}\sigma_{m',n'}^2+\mu'( \mu_{m',n'} - \mu'(\tau_0 + \tau_{m,m'}) )\sigma_{m,n}^2}{\mu^2\sigma_{m',n'}^2 +\mu'^2\sigma_{m,n}^2} \notag \\
& \sigma_G^2 = \frac{\sigma_{m,n}^2\sigma_{m',n'}^2}{\mu^2\sigma_{m',n'}^2 +\mu'^2\sigma_{m,n}^2}. \notag
\end{aligned}
\end{equation}
The expression in~\eqref{eq:output_matchedfilter_2} is equivalently written in a matrix form as
\begin{equation}
r(\tau_0,\mu,\mu') = \sigma_t\mathbf{s}^H\mathbf{R}(\tau_0,\mu,\mu')\mathbf{s},
\end{equation}
where $\mathbf{s} = \left [s_{1,1} \dots s_{N,1} , s_{1,2} \dots s_{N,2} \dots s_{N,M} \right]^T$ and $\mathbf{R}(\tau_0,\mu,\mu')$ contains the entries
\begin{equation}
\begin{aligned}
&R_{m,m',n,n'}(\tau_0,\mu,\mu') = \frac{\sigma_t\sqrt{\mu\mu'}e^{-j\omega_c\mu'(\tau_0 + \tau_{m,m'})}}{\sqrt{2\pi(\mu^2\sigma_{m',n'}^2 +\mu'^2\sigma_{m,n}^2)}}\cdot \\
& e^{-\frac{(\mu'\mu_{m,n}- \mu \mu_{m',n'} + \mu \mu'(\tau_0 + \tau_{m,m'}))^2}{2(\mu^2\sigma_{m',n'}^2 +\mu'^2\sigma_{m,n}^2)}} e^{\omega_c(\mu-\mu')(j\mu_G -\frac{\omega_c}{2}(\mu-\mu')\sigma_T^2)},
\end{aligned}
\end{equation}
in a proper order.

\subsection{Robust Waveform Design Based on Statistical Performance}
Target detection is formulated as a statistical problem as follows. Take a region $\mathcal{S}$ of the parameter pairs $(\mu,\tau)$ in which a good detection performance is desired. The detection is based on the output $r$ from a filter matched to a nominal point $(\tau',\mu')\in \mathcal{S}$. A matched filter\cite{Skolnik,Skolnik81} ensures maximal detection performance when the true parameters are nominal in the presence of white noise.

The idea is to develop a simple decision rule for $r$, which identifies whether a target is present in $\mathcal{S}$ or not. The problem can be formulated as a composite hypothesis testing, where $\mathcal{H}_0$ denotes the hypothesis that no source is present, in which $r$ is generated by a white Gaussian noise processes, and $\mathcal{H}_1$ denotes the composite hypothesis of source existence. Then, the likelihood functions under the different hypotheses are
\begin{equation}
\begin{aligned}
&\mathcal{H}_0:\ r\sim\mathcal{N}(0,\sigma^2\mathbf{s}^H\mathbf{R}_0\mathbf{s})\\
&\mathcal{H}_1:\ r~\sim\mathcal{N}(\sigma_t\mathbf{s}^H\mathbf{R}(\tau_0,\mu,\mu')\mathbf{s},\sigma^2\mathbf{s}^H\mathbf{R}_0\mathbf{s}),
\end{aligned}
\end{equation}
where $\mathbf{R}_0=\mathbf{R}(\tau_0=0,\mu',\mu')$ and $\sigma^2$ denotes the noise power. The pair $(\tau_0,\mu)$ denotes the unknown true parameters. Later, we simply refer to it as $\theta$.  Let us consider the Generalized Likelihood Ratio Test (GLRT)\cite{kay1998fundamentals}, which can be written as
\begin{equation}
\min\limits_{\theta,\sigma_t}|r-\sigma_t\mathbf{s}^H\mathbf{R}(\theta)\mathbf{s}|^2+\gamma\gtrless|r|^2,
\end{equation}
where the argument $\mu'$ is dropped as its value is assumed to be fixed, and $\gamma$ is a threshold. As $\sigma_t$ is free the minimization on the left hand side gives zero. Thus, the GLRT simplifies to
\begin{equation}
\label{eq:estimator}
\gamma\gtrless|r|,
\end{equation}
which is a simple power thresholding scheme.

For this detector probabilities of detection, $P_D$, and false alarm, $P_{FA}$, can be calculated to
\begin{equation}
\begin{aligned}
&P_D(\theta,\sigma_t,\gamma)=\frac{1}{\pi\mathbf{s}^H\mathbf{R}_0\mathbf{s}\sigma^2}
\int_{|r|>\gamma}e^{-\frac{|r-\sigma_t\mathbf{s}^H\mathbf{R}(\theta)\mathbf{s}|^2}{\mathbf{s}^H\mathbf{R}_0\mathbf{s}\sigma^2}}\text{d}\nu(r)\\
&P_{FA}=\frac{1}{\pi\mathbf{s}^H\mathbf{R}_0\mathbf{s}\sigma^2}
\int_{|r|>\gamma}e^{-\frac{|r|^2}{\mathbf{s}^H\mathbf{R}_0\mathbf{s}\sigma^2}}\text{d}\nu(r),
\end{aligned}
\end{equation}
where $\nu(\ldotp)$ denotes the Lebesgue measure on the complex plane of $r$.

To optimize the waveform, consider the worst detection performance $P_D(\theta,\sigma_t,\gamma)$ over all scenarios defined by $(\theta,\sigma_t)$, where $P_{FA}=\alpha$ is fixed. One may define the best design as the one maximizing the worst detection.
%which is mathematically expressed as
%\begin{equation}
%\begin{aligned}
%&\max\limits_{\mathbf{s}}\min\limits_{(\theta,\sigma_t)}&&P_D(\theta,\sigma_t,\gamma)\\
%&\text{s.t.}&&P_{FA}=\alpha,
%\end{aligned}
%\end{equation}
%where $\alpha$ is a design parameter.
Unfortunately, direct calculation shows that this approach fails in the current occasion as the worst detection performance is independent of the choice of waveform. To overcome this, denote by $P_{D,\text{worst}}(\alpha)$, or simply $P_{D,\text{worst}}$, the worst detection performance. Then, define an optimal design as follows:
\newtheorem{Definition}{Definition}
\newtheorem{theorem}{Theorem}

\begin{Definition}
\label{def}
A design $\mathbf{s}$ is optimal for a given value of $\alpha$ if for sufficiently small values of $\epsilon$, the set $\mathcal{S}_\epsilon$ of $\epsilon-$worse scenarios, defined by
\begin{equation}
\mathcal{S}_\epsilon=\{(\theta,\sigma_t)\mid|P_D(\theta,\sigma_t)<P_{D,\text{worst}}+\epsilon\}
\end{equation}
has a minimal area (Lebesgue measure) in any compact region.
\end{Definition}
Clearly, this gives a design, where it is least likely to encounter a low performance, although not impossible. The following theorem provides a practical method to realize this design.
\begin{theorem}
The optimal design in the sense of \textit{Definition~\ref{def}} is a solution to the following optimization
\begin{equation}\label{eq:opt}
\begin{aligned}
&\max_{\mathbf{s}} \min_{\theta} &&|\mathbf{s}^H\mathbf{R}(\theta)\mathbf{s}| \\
&\text {s.t.} &&\mathbf{s}^H\mathbf{R}_0\mathbf{s} = 1.
\end{aligned}
\end{equation}
\end{theorem}
The reader may notice that~\eqref{eq:opt} could be directly introduced and intuitively validated. It is simple to verify that $|\mathbf{s}^H\mathbf{R}(\theta)\mathbf{s}|$ and $\mathbf{s}^H\mathbf{R}_0\mathbf{s}$ are the share of signal and noise in the filter output energy over $\mathcal{S}$, respectively. Thus, \eqref{eq:opt} promotes a uniformly high signal-output energy. However, the above calculations tie \eqref{eq:opt} to a statistically sound detection approach.

Further, $\mathbf{R}_0$ is positive semi-definite, and the term $\mathbf{s}^H\mathbf{R}_0\mathbf{s}$ characterizes the energy of the signal, which is typical for a matched filter design. Thus, no extra care about the transmit energy is considered. Clearly, \eqref{eq:opt} guarantees high detection rate, but it does not consider false detection due to coupling between the filter and out-of-box sources. However, it is expected that the finite energy constraint automatically enforces low sidelobe energy. We later argue on the validity of this assumption by analyzing numerical results.

\section{Proposed Method for Solving \eqref{eq:opt}}
It is difficult to exactly solve \eqref{eq:opt}, as $\mathbf{R}(\theta)$ is in general not Hermitian. One approximate solution is to consider the inner optimization over a finite number of grid points, $\theta_1,\theta_2,\ldots,\theta_l$. Then, \eqref{eq:opt} can be written as
\begin{equation}\label{eq:opt2}
\begin{aligned}
&\max_{\mathbf{s}} \min_{\lambda_1,\lambda_2,\ldots,\lambda_l} &&\sum_k|\mathbf{s}^H\mathbf{R}(\theta_k)\mathbf{s}|\lambda_k \\
&\text {s.t.} &&\mathbf{s}^H\mathbf{R}_0\mathbf{s} = 1, \quad\sum_k\lambda_k=1,
\end{aligned}
\end{equation}
where $\lambda_k$ is a positive number. By changing the order of minimization and maximization in \eqref{eq:opt2} we obtain the following suboptimal design, which is simpler to solve \cite{bazaraa2013nonlinear}.
\begin{equation}\label{eq:opt_dual}
\begin{aligned}
& \min_{\lambda_1,\lambda_2,\ldots,\lambda_l} \max_{\mathbf{s}}&&\sum_k|\mathbf{s}^H\mathbf{R}(\theta_k)\mathbf{s}|\lambda_k \\
&\text {s.t.} &&\mathbf{s}^H\mathbf{R}_0\mathbf{s} = 1, \quad\sum_k\lambda_k=1.
\end{aligned}
\end{equation}
As opposed to the uniformly optimal design in the original order, the change of order results in an optimization considering average performance weighted by $\{\lambda_k\}$. However, as our simulation results indicate, the latter average design also leads to a remarkably good detection performance.
Now, the inner optimization in \eqref{eq:opt_dual} is simplified as follows. Consider the eigenvalue decomposition of $\mathbf{R}_0$, i.e.,
\begin{equation}
\mathbf{R}_0=\mathbf{U}\Sigma\mathbf{U}^H=\mathbf{U}_0\Sigma_0\mathbf{U}_0^H,
\end{equation}
where $\Sigma_0$ and $\mathbf{U}_0$ are the nonzero blocks of $\Sigma$ and its corresponding columns of $\mathbf{U}$, respectively. Let $\mathbf{u}=\mathbf{\Sigma_0}^\frac{1}{2}\mathbf{U_0}^H\mathbf{s}$, which implies that any vector $\mathbf{s}$ is uniquely decomposed in terms of its corresponding $\mathbf{u}$ as
\begin{equation}
\mathbf{s}=\mathbf{U_0}\mathbf{\Sigma_0}^{-\frac{1}{2}}\mathbf{u}+\mathbf{U_1}\mathbf{p},
\end{equation}
where $\mathbf{p}$ is a suitable vector and $\mathbf{U_1}$ spans the null space of $R_0$. Note that, any vector $\mathbf{s}$ with $\mathbf{s}^H\mathbf{R}_0\mathbf{s}=0$ corresponds to zero-energy, which leads to a zero output-signal. This clearly means that $\mathbf{R}(\theta)\mathbf{U}_1=0$ for every $\theta$. Thus, the term $U_1$ does not have any effect on the waveform design, and the inner optimization in \eqref{eq:opt_dual} can be expressed as
\begin{equation}\label{eq:opt_dual_3}
\begin{aligned}
& \max_{\mathbf{u}}  &&\sum_k\lambda_k | \mathbf{u}^H\tilde{\mathbf{R}}(\theta_k)\mathbf{u})| \\
&\text {s.t.} &&\|\mathbf{u}\|^2_2= 1,
\end{aligned}
\end{equation}
where $\tilde{\mathbf{R}}(\theta_k)=\mathbf{\Sigma}_0^{-1/2}\mathbf{U_0}^H\mathbf{R}(\theta_k)\mathbf{U_0}\mathbf{\Sigma_0}^{-\frac{1}{2}}$.

To solve \eqref{eq:opt_dual_3} we propose the following efficient scheme. First, note that, $|\alpha|=\max\limits_{\phi}\Re(e^{-j\phi}\alpha)$. Thus,  \eqref{eq:opt_dual_3} is equivalently written as
\begin{equation}\label{eq:cycle}
\begin{aligned}
&\max_{\mathbf{u},\phi_1,\phi_2,\ldots,\phi_l}\sum_k\lambda_k\Re(e^{-j\phi_k}\mathbf{u}^H\tilde{\mathbf{R}}(\theta_k)\mathbf{u})\\
&=\max_{\mathbf{u},\phi_1,\phi_2,\ldots,\phi_l}\mathbf{u}^H\mathbf{M}(\phi_1,\phi_2,\ldots,\phi_l)\mathbf{u},
\end{aligned}
\end{equation}
where
\small
\begin{equation}
\begin{aligned}
\mathbf{M}(\phi_1,\phi_2,\ldots,\phi_l)=\sum_k\lambda_k ( e^{-j\phi_k}\tilde{\mathbf{R}}(\theta_k) + e^{j\phi_k}\tilde{\mathbf{R}}^H(\theta_k)).
\end{aligned}
\end{equation}
\normalsize
As the optimization is performed over all unit vectors $\mathbf{u}$, the solution for a fixed choice of $\phi_1,\ldots,\phi_l$ is the eigenvector of $\mathbf{M}$ corresponding to the largest eigenvalue, which we denote by $\mathbf{u}_m(\mathbf{M})$ and $\lambda_m(\mathbf{M})$, respectively. Accordingly, we propose the following cyclic solution of the inner optimization.
\begin{enumerate}
\item Start from an arbitrary choice of $\phi_k^{0}$ and set $r=1$.
\item Get $\mathbf{M}^{r-1}=\mathbf{M}(\phi_1^{r-1},\ldots,\phi_n^{r-1})$ and set $\mathbf{u}^{r}=\mathbf{u}_m(\mathbf{M}^{r-1})$
%and $\lambda^{n}=\lambda(\phi^{n-1})$
by calculating $\lambda_m(\mathbf{M}(\phi^{r-1}))$.
\item Evaluate $\phi_k^{n}$ as the argument of the complex number $(\mathbf{u}^{r})^H\tilde{\mathbf{R}}(\theta_k)\mathbf{u}^{r}$ , update $r$ to $r+1$ and go to step 2.
\end{enumerate}
Step 2 and step 3 increase the cost of \eqref{eq:cycle} with respect to $\mathbf{u}$ and $\phi$. Thus, the cost monotonically increases, which guarantees convergence.

Once a solution, say $\bar{\mathbf{u}}$, for the inner optimization and given values of $\lambda_k$ is obtained, a local optimization technique such as steepest descent is performed to update $\lambda_k$ for the outer minimization. Although, the gradient, $\nabla F$, of the cost $F=F(\lambda_1,\ldots,\lambda_l)$ at a given point is simple to compute, applying a steepest decent technique is generally difficult.
%The gradient $\nabla F$ of the cost $F=F(\lambda_1,\ldots,\lambda_l)$ for the outer loop (the maximal value of the inner loop), at this given point is simple to compute.
%$\nabla F=\left[|\bar{\mathbf{u}}^H\tilde{\mathbf{R}}(\theta_k)\bar{\mathbf{u}}|\right],
%$
%\begin{equation}
%\nabla F=\left[|\bar{\mathbf{u}}^H\tilde{\mathbf{R}}(\theta_k)\bar{\mathbf{u}}|\right],
%\end{equation}
%where $[\cdot]$ denotes a vector for which its elements are given by the argument.
%The gradient is utilized in accordance with the constraint, which is difficult in general.
However, the number of grid points, $l$, in \eqref{eq:opt2} may be significantly low as a result of sufficient correlation between conceivable return signals. In fact, we employ only a pair of properly selected grid points, typically the corner points of a parameter box, from which the outer optimization can be substantially simplified to
%\begin{equation}
%\begin{aligned}
%&\min_{\lambda_1,\lambda_2}&&F(\lambda_1,\lambda_2)\\
%& \text{s.t.} && \lambda_1+\lambda_2=1,
%\end{aligned}
%\end{equation}
%which is equivalently written as
\begin{equation}
\begin{aligned}
&\min_{0\leq \lambda\leq 1}F(\lambda,1-\lambda),\\
\end{aligned}
\end{equation}
and a resulting 1-dimensional optimization is either carried out by a grid search or a bisection method\cite{fletcher13}.% Still, numerical results imply that such a design provides a good performance.

\section{Numerical Validation}
For different choices of regions in which a robust performance is desired, the efficiency of the proposed algorithm is presented in terms of average and minimum correlation as well as Receiver Operating Characteristic (ROC)\cite{richards05} curves. To examine the method, we calculate its efficiency by Monte-Carlo simulation in selected scenarios. The number of waveform generators is $M=3$, for each $N=30$ basis functions are generated. The system has a bandwidth-time product of $200$, the nominal value of the time-scaling is $\mu' = 0.94$, and the target's reflection coefficient is set to $\sigma_t =1$.

The Gaussian basis functions are generated with mean values, $\mu_k$, uniformly located over the pulse duration, and their standard deviations are selected randomly within an interval $[\sigma_{\min,k}\ \sigma_{\max,k}]$, where $\sigma_{\max,k}$ is such that the effective interval between $\mu_k\pm 3\sigma_k$ is ensured to entirely lie within the pulse time, and $\sigma_{\min,k}$ restricts the highest effective frequency component.

The region in which the performance is evaluated is chosen as a box, which varies in size. The smallest box is within the confines of $\mu \in [\mu_0-\epsilon_{\mu},\mu_0+\epsilon_{\mu}]$ and $\tau \in [-\epsilon_{\tau},\epsilon_{\tau}]$, where $\epsilon_{\mu}$ and $\epsilon_{\tau}$ are determined with respect to twice the system's resolution limits. In order to investigate the effect of varying uncertainty, the box size increases with a factor $\beta$ to $\epsilon_{\mu}/\beta$ and $\epsilon_{\tau}/\beta$, where $\beta = [0.9,\dots,0.1]$.% The waveforms are then optimized to have good performance for the different boxes.

The minimum and average in-box correlation, $|\mathbf{s}^H\mathbf{R}(\theta)\mathbf{s}|$, are presented in Table~\ref{tab:correlation_avg} and Table~\ref{tab:correlation_min}, in which the results are averaged over $100$ random choices of basis functions. The correlation properties are evaluated over a dense grid and then taking average or minimum, respectively. The outcome is compared with the cases where a Linear Frequency Modulated (LFM) pulse and a single Gaussian pulse is transmitted from each signal generator, with the same bandwidth-time product. It should be noted that the LFM pulse provides a high resolution, and is not expected to give a robust performance. %The outcome is compared with the case where a single Gaussian pulse is transmitted from each signal generator.
\begin{table}[h!]
\caption{Average correlation properties of the designed waveforms.}
\label{tab:correlation_avg}
\begin{tabular}{l c c c c c}
Algorithm & $\beta =1$ & $\beta =0.8$& $\beta =0.6$ & $\beta =0.4$& $\beta =0.2$\\
 \hline
 \textit{Proposed alg.} & $0.99$ &  $0.98$ & $0.97$ & $0.94$ & $0.65$ \\
 \textit{Gaussian} & $0.97$ &  $0.93$ & $0.88$ & $0.80$ & $0.60$ \\
   \textit{LFM} & $0.30$ &  $0.26$ & $0.21$ & $0.15$ & $0.09$
\end{tabular}
\end{table}

\begin{table}[h1]
\caption{Minimum correlation properties of the designed waveforms.}
\label{tab:correlation_min}
\begin{tabular}{l c c c c c}
Algorithm & $\beta =1$ & $\beta =0.8$& $\beta =0.6$ & $\beta =0.4$& $\beta =0.2$\\
 \hline
 \textit{Proposed alg.} & $0.98$ &  $0.96$ & $0.93$ & $0.85$ & $0.41$ \\
 \textit{Gaussian} & $0.88$ &  $0.83$ & $0.70$ & $0.42$ & $0.1$ \\
  \textit{LFM} & $0.0039$ &  $0.0018$ & $0.0018$ & $0.0018$ & $0.0016$
\end{tabular}
\end{table}

\begin{figure}[h!]
\centering
\includegraphics[width=8.5cm]{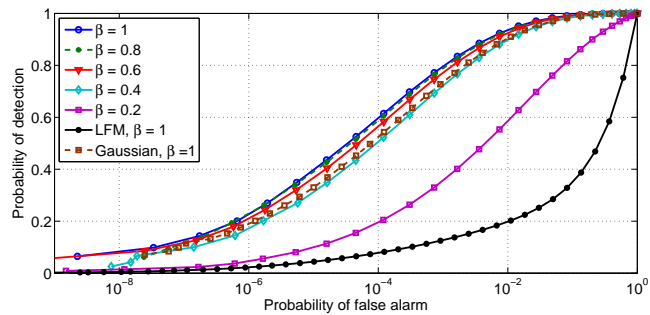}
\caption{ROC curves when the SNR is $10$~dB for regions specified by $\beta = [1,0.8,0.6,0.4,0.2]$.}
\label{fig:ROC_curve}
\end{figure}

Figure~\ref{fig:ROC_curve} shows the ROC curves. These curves are calculated from the statistical formulation through a Monte Carlo simulation using the estimator in \eqref{eq:estimator} with $10^6$ noise realizations, and a Signal-to-Noise Ratio (SNR) of $10$~dB. The curves are also averaged over $10$ independent draws of basis functions. The Figure shows the characteristics when $\beta = [1,0.8,0.6,0.4,0.2]$, for a varying threshold $\gamma = [0,0.05,\dots,4]$.

The curve corresponding to $\beta=1$ illustrates the performance in which the smallest box is selected. This curve is the nominal performance. As seen, the performance decreases when expanding the region. However, it exhibits a robust behavior for relatively large boxes.

\begin{figure}[h!]
\centering
\includegraphics[width=8.5cm]{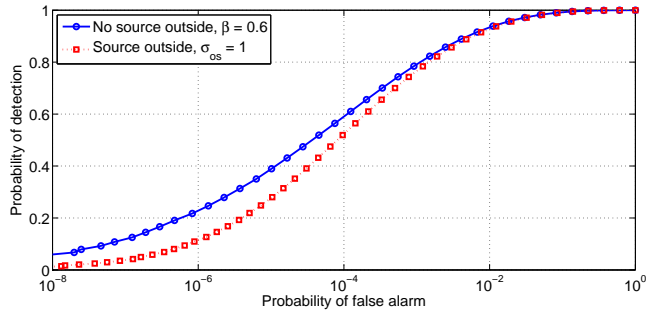}
\caption{ROC curve when an out-of-region source with $\sigma_{os} = 1$ is present. The SNR is $10$~dB and $\beta = 0.6$.}
\label{fig:ROC_curve_outsource}
\end{figure}

The last part of the numerical results presents how an out-of-region source will affect the ROC curve. This kind of source increases the probability of false alarm if its return is highly correlated with the matched filter for the region of interest. The position of the source is randomly generated outside the interval of $\mu$ and $\tau$. Results shown in Figure~\ref{fig:ROC_curve_outsource} illustrates ROC curves when the out-of-region source has a reflection coefficient of $\sigma_{os} =1$. The outcome when $\sigma_{os} =-1$ coincides with $\sigma_{os}=1$, and is therefore omitted in the Figure. The result is compared with the corresponding ROC curve when no such source exists for waveforms optimized with $\beta =0.6$.

\section{Conclusions}
For a wideband radar, we considered a robust technique for waveform design within relatively wide parameter ranges. The method was developed from a statistical framework for detecting a single target, and simplified by approximation to obtain a tractable design. Being robust implies that the waveforms are not designed to provide good resolution properties. Therefore, a next step is to apply a super-resolution technique to the restricted area obtained from this first stage.

The correlating filters at the receiver side were selected to have a matched filter structure. We remark that, a different kind of filter design, e.g., \cite{Ackroyd73,Stoica2008,Stoica2008_2} might result in better performance, which is a topic for future investigation.

As shown by numerical validation, the method ensured reliable detection in the desired range of target parameters. The probabilities of detection and false alarm are illustrated with ROC curves, which showed a small loss of performance when increasing the desired region of reliable detection up to a certain size. %The sudden loss of performance, for a large region, might be related to the insufficient number of grid points in the optimization. However, this is a subject of further investigation.
The outcome was compared with two conventional transmit signal designs and showed an increased performance for the investigated problem. Note that, LFM signals have good resolution properties, which makes them unsuitable for the discussed application. It was also seen that in presence of an out-of-region source the detection properties are only slightly affected, which implied that the design promoted a low sidelobe energy.

\bibliographystyle{IEEEtran}
\bibliography{References}
\end{document}